# Unrealizable Learning in Binary Feed-Forward Neural Networks

**M. Sporre**
Nordita
Blegdamsvej 17
DK-2100 Copenhagen Ø
Denmark
sporre@nordita.dk

**ABSTRACT**

Statistical mechanics is used to study unrealizable generalization in two large feed-forward neural networks with binary weights and output, a perceptron and a tree committee machine. The student is trained by a teacher being larger, i.e. having more units than the student. It is shown that this is the same as using training data corrupted by Gaussian noise. Each machine is considered in the high temperature limit and in the replica symmetric approximation as well as for one step of replica symmetry breaking. For the perceptron a phase transition is found for low noise. However the transition is not to optimal learning. If the noise is increased the transition disappears. In both cases $\epsilon_g$ will approach optimal performance with a $(\ln\alpha/\alpha)^k$ decay for large $\alpha$. For the tree committee machine noise in the input layer is studied, as well as noise in the hidden layer. If there is no noise in the input layer there is, in the case of one step of replica symmetry breaking, a phase transition to optimal learning at some finite $\alpha$ for all levels of noise in the hidden layer. When noise is added to the input layer the generalization behavior is similar to that of the perceptron. For one step of replica symmetry breaking, in the realizable limit, the values of the spinodal points found in this paper disagree with previously reported estimates [1],[2]. Here the value $\alpha_{sp} = 2.79$ is found for the tree committee machine and $\alpha_{sp} = 1.67$ for the perceptron.

## 1. Introduction

A Feed-forward neural network can be used to estimate an unknown rule from random examples [3] by adaption of its weights. Using methods from statistical mechanics of disordered systems [4] the performance of a student network trained on examples obtained from a teacher network of the same architecture has been studied (for a review see [5]). In this case the rule is said to be realizable since it is possible for the student to develop the same weights as the teacher.

One way to construct an unrealizable rule is to allow for a teacher that is larger (more units) than the student. This will be shown to be equivalent to adding Gaussian noise to the training set. The noisy data scenario has been investigated for networks with continuous weights [6],[7],[8]. In the limit where the teacher is infinitely larger than the student (large noise limit) the only thing the student can do is to learn each example by heart, and in this limit the problem reduces to that of storage capacity.

In this paper the generalization behavior of two different types of binary neural networks with binary weights is studied, a perceptron (section 2) and a tree committee machine (section 3), in the limit where the number of units is large. The rule is defined by a teacher of the same type but having more units than the student, making the task unrealizable.

The training of the student, having $N$ units, is based on $\alpha N$ examples obtained by picking inputs $\vec{\xi}^\mu$ and assigning outputs $\tau^\mu$ as given by the teacher. With $\sigma^\mu$ being the $\mu$th output of the student, a training energy $E = \sum_\mu \theta(-\sigma^\mu \tau^\mu)$ is defined, which leads to a probability density with Boltzmann weight $e^{-\beta E}$, where $\beta = 1/T$ is the inverse temperature. First the high temperature limit is considered for each type of network, the perceptron in section (2.1) and the tree committee machine in section (3.1). Then, in sections (2.2) and (3.2), the replica trick is used, assuming replica symmetry (RS), to study the average over all training sets of the free energy, $\beta f$. In sections (2.3) and (3.3) the corrections given by one step of replica symmetry breaking (RSB) are discussed. Since, in the noiseless limit, the value of the spinodal point found in the RSB-case disagrees with previously reported estimates [1],[2], some time is spent on the saddle point equations in appendix A. Finally, in appendix B, the procedure for finding the asymptotic generalization behavior for large $\alpha$ is given.

## 2. A Large Binary Perceptron.

Let the student and the teacher have $N$ and $M$ input units respectively with $N \leq M$. Presented an input, $\vec{s}$, the teacher evaluates, $\tau(\vec{s}) = \text{sgn}(\vec{v} \cdot \vec{s})$, while the student computes, $\sigma(\vec{s}_0) = \text{sgn}(\vec{w}_0 \cdot \vec{s}_0)$, given the input $\vec{s}_0$. Here $\vec{s}$ and $\vec{v}$ are elements of $\mathcal{R}^M$, while vectors having a zero subscript are elements of $\mathcal{R}^N$. When the student is presented the same input vector, $\vec{\xi}$, as the teacher, it only considers the $N$ first components, $\vec{\xi}_0$. Thus the target rule will be,

$$\begin{aligned}
\tau(\vec{\xi}) &= \text{sgn}\left[\frac{1}{\sqrt{M}}\sum_{j=1}^N v_j \xi_j + \frac{1}{\sqrt{M}}\sum_{j=N+1}^M v_j \xi_j\right] \\
&= \text{sgn}\left[\sqrt{\frac{N}{M}}\left(\vec{v}_0 \cdot \vec{\xi}_0 + \frac{1}{\sqrt{N}}\sum_{j=N+1}^M v_j \xi_j\right)\right] \\
&= \text{sgn}\left(\vec{v}_0 \cdot \vec{\xi}_0 + \eta\right) , \quad (2.1)
\end{aligned}$$

where $\vec{v}_0$ is constructed from the first $N$ components of $\vec{v}$. Effectively this means that the student will be given the task $\tau'(\vec{\xi}_0) = \text{sgn}(\vec{v}_0 \cdot \vec{\xi}_0)$ with noise on the input vector $\vec{\zeta}_0 = \vec{\xi}_0 + \vec{\kappa}_0$ and/or on the weight vector $\vec{J}_0 = \vec{v}_0 + \vec{\omega}_0$. Since $\eta$ is constructed from independent Gaussian random variables, $v_j$ and $\xi_j$ ($j = N+1, ..., M$) with unit variance, $\eta$ will also be Gaussian with variance,

$$\begin{aligned}
\langle \eta^2 \rangle &= \left\langle \frac{M-N}{N}\left(\frac{1}{\sqrt{M-N}}\sum_{j=N+1}^M v_j \xi_j\right)^2 \right\rangle \\
&= \frac{1-\gamma^2}{\gamma^2} , \quad (2.2)
\end{aligned}$$

where $\gamma = \sqrt{N/M}$. $\gamma$ has the simple interpretation of the relative size of the student to the teacher. If $\gamma = 1$ the student and the teacher are of the same size, i.e. there is no noise. If $\gamma = 0$ the teacher is infinitely larger than the student, i.e. the data will be completely noisy.

The generalization error, $\epsilon_g$, obtained by taking the average of $\theta(-\sigma\tau)$ over normal distributed inputs, $s_j$ ($j = 1, ..., M$), is

$$\epsilon_g = \frac{1}{\pi}\arccos(\gamma R) , \quad (2.3)$$

where $R$ is the overlap between $\vec{w}_0$ and $\vec{v}_0$. For $R = 1$ we obtain the optimal value, $\epsilon_{opt}$, of $\epsilon_g$.

First the high temperature limit is considered. Then by using the replica method, the RS approximation is studied, and finally the corrections given by one-step RSB are discussed.

### 2.1. High Temperature Limit

In previous work [1] the high temperature limit has proven to be interesting since it is both computationally easy and gives the general behavior of learning. It is defined so that both $\alpha$ and $T$ approach infinity while $\alpha\beta$ remains constant. The free energy is simply

$$\begin{aligned}
\beta f &= \frac{1-R}{2}\ln(\frac{1-R}{2}) + \frac{1+R}{2}\ln(\frac{1+R}{2}) \\
&\quad + \frac{\alpha\beta}{\pi}\arccos(\gamma R) . \quad (2.1.1)
\end{aligned}$$

The qualitative behavior of the learning curves can be divided into two types depending on whether the noise level is above or below a particular value $\gamma_0$. For $\gamma_0 < \gamma < 1$ there is, as in the realizable case, a range $(\alpha\beta)_{sp1} < \alpha\beta < (\alpha\beta)_{sp2}$, for which $\beta f$ has two minima. In between $(\alpha\beta)_{sp1}$ and $(\alpha\beta)_{sp2}$ there is a transition point $(\alpha\beta)_{tr}$ at which the global properties of the minima change. In contrast to the noiseless case, $(\alpha\beta)_{sp1} > 0$ and thus for $0 \leq \alpha\beta \leq (\alpha\beta)_{sp1}$ there is only one minimum. The minimum persisting also for $\alpha\beta \geq (\alpha\beta)_{sp2}$ is close to $R = 1$ and approaches optimal performance as $\alpha\beta$ increase. Note that in contrast to the realizable case there is no solution at $R = 1$. Typically $(\alpha\beta)_{tr}$, $(\alpha\beta)_{sp1}$ and $(\alpha\beta)_{sp2}$ increase with decreasing $\gamma$ and merge at $\gamma = \gamma_0$. This is illustrated in figure (1).

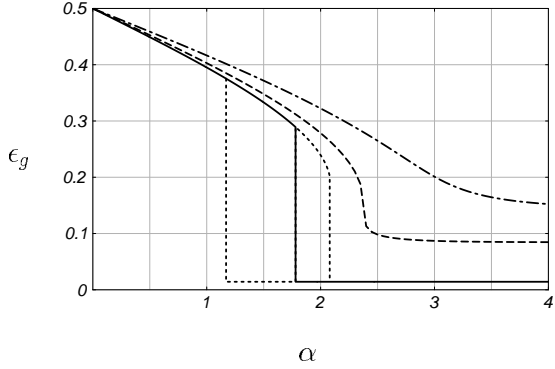

Figure 1: *The picture show learning curves for the perceptron in the high temperature limit at three different noise levels. At $\gamma = 0.999$ there are two states a stable (solid line) and a metastable (dotted) one resulting in a phase transition. At $\gamma = 0.965$ (dashed) the spinodal points and the transition point merge and finally at $\gamma = 0.9$ (dashed-dotted) there is no phase transition.*

The two minima of $\beta f$ must be separated by a maximum, implying that $\frac{\partial^2 \beta f}{\partial R^2} = 0$ at the spinodal points. Using the saddle point equation,

$$\frac{\alpha\beta\gamma}{\pi\sqrt{1-\gamma^2 R^2}} = \frac{1}{2}\ln\left(\frac{1+R}{1-R}\right) = \operatorname{arctanh}(R), \quad (2.1.2)$$

to eliminate $\alpha\beta$ from $\frac{\partial^2 \beta f}{\partial R^2} = 0$, gives

$$R = \tanh\left[\frac{1-\gamma^2 R^2}{\gamma^2 R(1-R^2)}\right] \equiv g(R,\gamma). \quad (2.1.3)$$

For $\gamma = 1$, (2.1.3) has one solution, $R_{sp} = 0.83$ resulting in $(\alpha\beta)_{sp} = 2.08$ in agreement with [1]. In the region $\gamma_0 < \gamma < 1$ (2.1.3) has two solutions giving $(\alpha\beta)_{sp1}$ and $(\alpha\beta)_{sp2}$. At $\gamma = \gamma_0$ the two solutions merge and the two curves $R$ and $g(R,\gamma)$ are tangent to each other. Thus $\gamma_0$ can be found by solving,

$$\frac{\partial}{\partial R} g(R,\gamma_0) = 1, \quad (2.1.4)$$
$$g(R,\gamma_0) = R, \quad (2.1.5)$$

giving $\gamma_0 = 0.965$. For $\gamma < \gamma_0$, $\beta f$ has only one minimum (for all $\alpha\beta$) which moves towards $R = 1$ as $\alpha\beta$ approaches infinity. Note that fairly small amounts of noise will change the qualitative behavior from phase transition to no transition.

In weight space this behavior can be understood as follows. In the noiseless case there are, for small $\alpha$, two regions in weight space corresponding to the minima of $\beta f$, one with poor generalization and one with good. If $\alpha$ is small enough the "poor" region has the lowest free energy. As $\alpha$ increase the "poor" region moves towards the "good" and for $\alpha > \alpha_{tr}$ the "good" region has the lowest free energy. Since for $\alpha = \alpha_{tr}$ the "poor" and "good" regions are separated, there will be a phase transition.

If noise is added, the sizes of these regions will increase. For low $\alpha$ there is only one region in weight space corresponding to a minimum of $\beta f$. It will have poor generalization. At $\alpha = \alpha_{sp1}$ another region corresponding to a free energy minimum appears. This region gives better generalization. Again as $\alpha$ increase the "poor" region moves towards the "good" and for $\alpha > \alpha_{tr}$ the "good" region has the lowest free energy. Since for $\alpha = \alpha_{tr}$ the "poor" and "good" regions are separated there will be a phase transition. If the noise is increased the "poor" region is so large that when the "good" region is created it will overlap with the "poor". Thus there is only one region, moving towards better generalization and there is no phase transition.

## 2.2. Replica Symmetric Theory

Using the same methods as in [1] the RS approximation to the free energy is obtained,

$$\beta f_{RS} = \underset{R,\hat{R},q,\hat{q}}{extr} \left[ G_r(R,q,\alpha,\gamma,\beta) + G_s(R,\hat{R},q,\hat{q}) \right],$$

$$G_r = -2\alpha \int Du\, H\left[\frac{\gamma R u}{\sqrt{q-\gamma^2 R^2}}\right] V\left(u\sqrt{\frac{q}{1-q}}\right),$$

$$G_s = \frac{1}{2}(1-q)\hat{q} + R\hat{R}$$
$$\quad - \int Du\, \ln\left[2\cosh\left(\hat{R} + \sqrt{\hat{q}}u\right)\right],$$

$$V(x) = \ln\left[e^\beta + \left(1-e^{-\beta}\right) H(x)\right]. \quad (2.2.1)$$

The saddle point equations generated by the extremal condition in (2.2.1) is given in appendix A. Here $q$ is the typical overlap between two different $\vec{w}_0$. $R$, $\gamma$ and $\alpha$ have the interpretation given above.

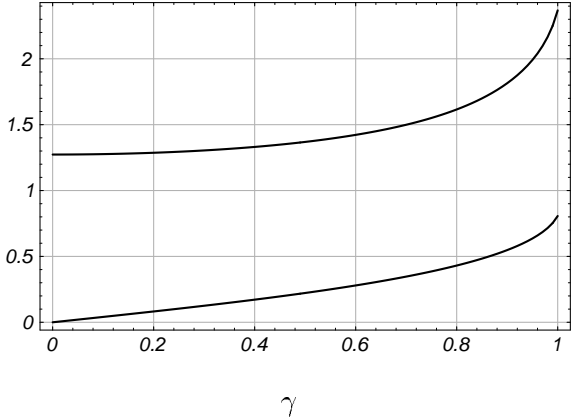

Figure 2: *Critical capacity for the perceptron in the RS-approximation. The upper curve shows $\alpha_c$ as a function of noise and the lower shows $R_c$.*

Using the saddle point equations we can, given $\gamma$ and $\beta$, eliminate the auxiliary variables $\hat{R}$ and $\hat{q}$, and find the dependence of $R$ (and $\epsilon_g$) on $\alpha$.

First consider the zero temperature case. This corresponds to only allowing students that answers all questions correctly. If $\gamma < 1$ the training data is noisy and there is a maximum size of the training set $\alpha_c N$ beyond which no student can perform optimally. $\alpha_c(\gamma)$ and $R_c(\gamma)$ are plotted in figure (2). For $\gamma = 0$ the known result of Gardner [9] is reproduced. Note that the curves do not give $\alpha_c \to \infty$ as $\gamma \to 1$. However this may not be expected since the curves only give correct predictions for states that are stationary points of $\beta f_{RS}$ and in the realizable case the state $R = 1$ is not stationary as was shown in [1]. For $\gamma = 1$ both the transition and the spinodal points agrees with the values found in [1]. A learning curve for $\gamma = 0.99$ is shown in figure (3).

At $T > 0$ the learning behavior is the same as for the high temperature limit but with a different $\gamma_0$, depending on $T$, and with $\epsilon_g$ and $q$ having the asymptotic form,

$$\epsilon_g - \epsilon_{opt} = C_1(\gamma, \beta) \frac{\ln \alpha}{\alpha} , \quad (2.2.2)$$

$$1 - q = C_(\gamma, \beta) \left(\frac{\ln \alpha}{\alpha}\right)^2 , \quad (2.2.3)$$

for large $\alpha$. For details on how to compute the asymptotic form see appendix B. For some range of $\gamma$, $\gamma_A < \gamma < 1$, there is a phase transition already at zero temperature while in a range $\gamma_B < \gamma < \gamma_A$ there is no transition at low temperature. As the temperature is increased a transition develops which is illustrated in figure (3) for $\gamma = 0.99$. Finally when $\gamma < \gamma_B$ there seems to be no phase transition no matter how high the temperature.

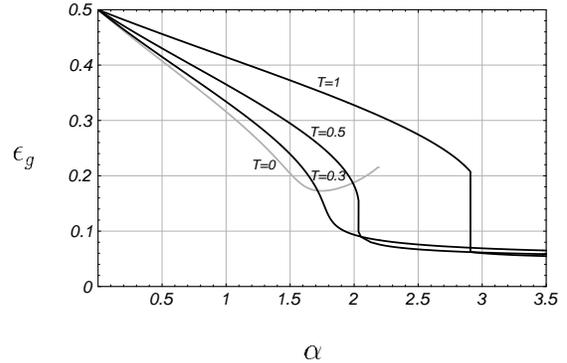

Figure 3: *For the RS-estimate of the perceptron the figure shows learning curves at the indicated temperature (T = 0, 0.3, 0.5, 1) for $\gamma = 0.99$. Only the stable state is plotted at $T = 0.5$ and $T = 1$. Note that for $T = 0$ there is a critical $\alpha$ above which the saddle point equations does not have a solution.*

## 2.3. Replica Symmetry Breaking

In the RS approximation the entropy will always turn negative at some finite $\alpha$ and therefore a region in $\alpha T$-space for which the system exhibits replica symmetry breaking (RSB) is expected, see figure (4). Analogous to [1] one step of RSB gives,

$$\beta f_{RSB} = extr \left[G_r(R, q_0, q_1, m, \alpha, \gamma, \beta) \right.$$
$$\left. + G_s(R, \hat{R}, q_0, \hat{q}_0, q_1, \hat{q}_1, m)\right] ,$$

$$G_r = -\frac{2\alpha}{m} \int Dt \int_0^\infty D\mu$$
$$\times \ln\left[\int D\omega \left(e^{-\beta} + \left(1 - e^{-\beta}\right) H(z)\right)^m\right] ,$$

$$G_s = \frac{1}{2}\left((m-1)q_1 + 1\right)\hat{q}_1 - \frac{m}{2}q_0\hat{q}_0 + R\hat{R}$$
$$+ \frac{1}{m} \int Dt \ln\left[\int Dy \left(2\cosh\left(\hat{R} + \sqrt{\hat{q}_0}t + \sqrt{\hat{q}_1 - \hat{q}_0}y\right)\right)^m\right] ,$$

$$z \equiv \frac{t\sqrt{q_0 - \gamma^2 R^2} - \mu\gamma R + \omega\sqrt{q_1 - q_0}}{\sqrt{1 - q_1}} , \quad (2.3.1)$$

where the extremum is taken over $R$, $\hat{R}$, $q_0$, $\hat{q}_0$, $q_1$,

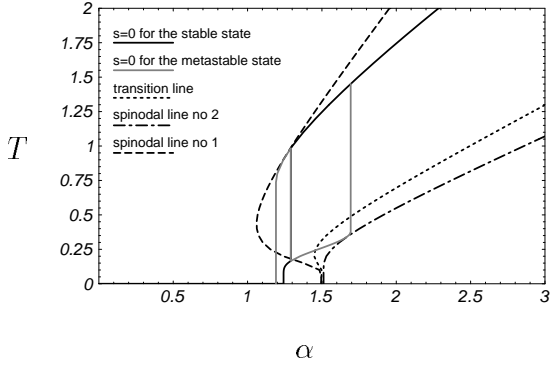

Figure 4: *Phase diagram for the perceptron at $\gamma = 0.999$. Below the solid line the stable state exhibits RSB and below the gray line the metastable line exhibits RSB. To the left of the first spinodal line there is only one state in the interior. Below the second spinodal line there is only the state close to $R = 1$. In between the two states coexist. Their global properties changes as the transition line is crossed.*

$\hat{q}_1$, and $m$. As in [1] the limit $q_1 \to 1$, $\hat{q}_1 \to \infty$ is considered, implying that the stationary points of $f_{RSB}$ are given by the stationary points of $f_{RS}$ having zero entropy (see appendix A for details).

The learning behavior is analogous to the high temperature limit but with $\gamma_0 = 0.995$. In appendix B the asymptotic form of $\epsilon_g$ and $q$ is computed,

$$\epsilon_g - \epsilon_{opt} = C_3(\gamma) \left(\frac{\ln \alpha}{\alpha}\right)^2, \quad (2.3.2)$$

$$1 - q = C_4(\gamma) \left(\frac{\ln \alpha}{\alpha}\right)^2. \quad (2.3.3)$$

When $0.999 \leq \gamma < 1$, $\alpha_{tr}$ occurs in between $\alpha_{sp1}$ and $\alpha_{sp2}$ while for $0.996 \leq \gamma \leq 0.998$, $\alpha_{tr} = \alpha_{sp1}$, i.e. the state with better generalization is stable as soon as it appears.

For the case $\gamma = 0.05$ the critical capacity, $\alpha_c = 0.83$ ($q_c = 0.56$) is found which is compatible with the known results for $\gamma = 0$ [10]. Some values are given in table (1) and some typical learning curves are given in figure (5).

In the noiseless limit the result, $\alpha_{sp} = 1.67$, correcting a previous result by Seung *et. al.* [1] ($\alpha_{sp} = 1.63$). The reason for this is given in appendix A.

It is also interesting to compare with some recently reported upper bounds for the Ising perceptron [11]. In this article the asymptotical behavior was found

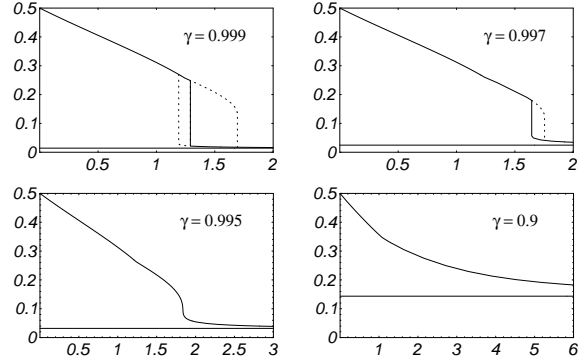

Figure 5: *Learning curves ($\epsilon_g$ vs $\alpha$) for the perceptron at the indicated noise levels. The solid line is the stable state, the dotted the metastable state, and the lower line indicates optimal performance.*

| $\gamma$ | $\alpha_{sp1}$ | $\alpha_{tr}$ | $\alpha_{sp2}$ |
|---|---|---|---|
| 1.0000 | 0.00 | 1.24 | 1.67 |
| 0.9999 | 0.53 | 1.25 | 1.67 |
| 0.9990 | 1.19 | 1.29 | 1.69 |
| 0.9970 | 1.64 | 1.64 | 1.75 |
| 0.9952 | 1.83 | 1.83 | 1.83 |

Table 1: *For one step of RSB for the perceptron the table shows some values of $\alpha_{tr}$, $\alpha_{sp1}$ and $\alpha_{sp2}$ for different $\gamma$*

to be the same as (2.3.2). The authors found that the phase transition disappeared below $\gamma = 0.998$ thus not only predicting the correct qualitative behavior but also giving a tight quantitative bound on $\gamma_0$. Also, at $\gamma = 0.998$, they found $\alpha_{tr} = 2.6136$ whereas $\alpha_{tr} = 1.83$ is obtained at $\gamma_0$ given above and using the replica method.

## 3. A Binary Tree Committee Machine.

Let the student and the teacher have $N$ ($K$) and $M$ ($L$) input (hidden) units respectively, with $N \leq M$, and $K \leq L$. We can think of the student (teacher) as a committee of binary perceptrons each of which has $N/K$ ($M/L$) input units. As the $l$th perceptron in the teacher is presented an input $\vec{s}_l$ the teacher evaluates,

$$\tau(\vec{s}_1, ..., \vec{s}_L) = \mathrm{sgn}\left[\frac{1}{\sqrt{L}} \sum_{l=1}^{L} \mathrm{sgn}\left(\sqrt{\frac{L}{M}} \sum_{m=1}^{M/L} v_{lm} s_{lm}\right)\right] , \quad (3.1)$$

while the student computes,

$$\sigma(\vec{s}_1^{(0)}, ..., \vec{s}_K^{(0)}) = \mathrm{sgn}\left[\frac{1}{\sqrt{K}} \sum_{k=1}^{K} \mathrm{sgn}\left(\sqrt{\frac{K}{N}} \sum_{m=1}^{N/K} w_{kn}^{(0)} s_{kn}^{(0)}\right)\right] , (3.2)$$

as the $k$th perceptron in the student is given the input $\vec{s}_k^{(0)}$. Here $\vec{s}_l$ and $\vec{v}_l$ are elements of $\mathcal{R}^{M/L}$ whereas a zero superscript indicates that the vector is an element of $\mathcal{R}^{N/K}$. When the student is presented the same set of input vectors, $\vec{\xi}_l$ ($l = 1, ..., L$), as the teacher it only considers the first $N/K$ components of the first $K$ vectors in that set, $\vec{\xi}_l^{(0)}$ ($l = 1, ..., K$). Analogous to the simple perceptron we find that this is equivalent to learning a noisy target rule,

$$\tau(\vec{\xi}_1^{(0)}, ..., \vec{\xi}_K^{(0)}) = \mathrm{sgn}\left[\frac{1}{\sqrt{K}} \sum_{k=1}^{K} \mathrm{sgn}\left(\sqrt{\frac{K}{N}} \sum_{n=1}^{N/K} v_{lm}^{(0)} \xi_{lm}^{(0)} + \eta_k\right) + \eta\right] , \quad (3.3)$$

where $\eta$ and $\eta_k$ are independent Gaussian random variables with variance,

$$\langle \eta^2 \rangle = \frac{L-K}{K} \equiv \frac{1-\gamma^2}{\gamma^2} , \quad (3.4)$$

$$\langle \eta_k^2 \rangle = \frac{KM}{NL} \equiv \frac{1-\delta^2}{\delta^2} . \quad (3.5)$$

$\gamma = \sqrt{K/L}$ is simply the relative number of hidden units of the student to the teacher while $\delta = \sqrt{NL/(KM)}$ is the relative number of input units of a perceptron in the student committee to a perceptron in the teacher committee. Thus $\gamma$ quantifies the noise in the hidden layer and $\delta$ the noise in the input layer. If $\gamma = \delta = 1$ the realizable case is recovered.

Using these parameters the generalization error is found,

$$\epsilon_g = \frac{1}{\pi} \arccos[R_e] , \quad (3.6)$$

where the effective order parameter is given by $R_e = \frac{2}{\pi} \gamma \arcsin(\delta R)$ and $R$ is the typical overlap between $\vec{w}_k^{(0)}$ and $\vec{v}_k^{(0)}$. Here, analogous to Schwarze and Hertz [2], it is assumed that $R$ is independent of the hidden unit index $k$.

As for the perceptron case the high temperature limit is considered first. Then by using the replica method, the RS approximation is studied and finally the corrections given by one step of RSB are discussed.

### 3.1. High Temperature Limit.

Taking the limits $T \to \infty$ and $\alpha \to \infty$ while keeping $\alpha\beta$ fixed the free energy is found,

$$\beta f = \frac{1-R}{2}\ln(\frac{1-R}{2}) + \frac{1+R}{2}\ln(\frac{1+R}{2}) + \frac{\alpha\beta}{\pi}\arccos(R_e) . \quad (3.1.1)$$

If the noise level is low enough, there exists two spinodal points, $\alpha_{sp1}$ and $\alpha_{sp2}$, with a phase transition in between. In contrast to the perceptron one find that if there is no input noise ($\delta = 1$) there is a phase transition to optimal performance at a finite $\alpha$ for all values of $\gamma$. Given a $\gamma$ and that $\delta_0(\gamma) < \delta$ a transition to a state approaching optimal learning in the large $\alpha$ limit is found. For $\delta < \delta_0(\gamma)$ the transition vanishes and $\epsilon_g$ approaches $\epsilon_{opt}$ as $\alpha$ tends to infinity. Especially if $\delta > \delta_A = \delta_0(0)$ there is always a phase transition while for $\delta < \delta_B = \delta_0(1)$ there is no phase transition independent of the hidden noise. By the same procedure as in section (2.1) one find $\delta_A = 0.965$, $\delta_B = 0.924$ and $\delta_0(\gamma)$ as shown in figure (6). Also here $\alpha_{sp1}$, $\alpha_{tr}$ and $\alpha_{sp2}$ increase with increasing noise.

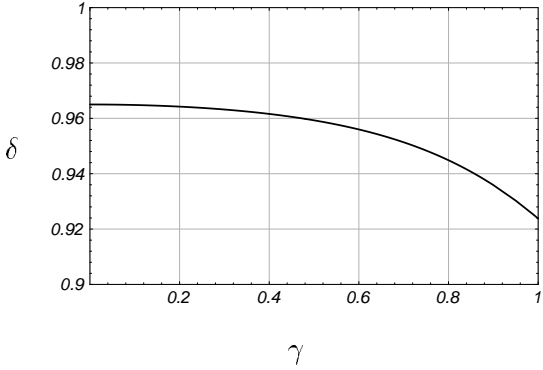

Figure 6: *In the high temperature limit for the tree committee machine the system undergoes a phase transition at some $\alpha_{tr}$ for noise levels above the curve. Below it the system will smoothly approach $\epsilon_{opt}$ as $\alpha \to \infty$.*

## 3.2. Replica Symmetric Theory.

Analogous to Schwarze and Hertz [2] the RS estimate to the free energy is found,

$$\beta f_{RS} = \underset{R,\hat{R},q,\hat{q}}{extr} \left[ G_r(R,q,\alpha,\gamma,\beta) + G_s(R,\hat{R},q,\hat{q}) \right] ,$$

$$G_r = -2\alpha \int Du \; H\left[\frac{R_e u}{\sqrt{q_e - R_e^2}}\right] V\left(u\sqrt{\frac{q_e}{1-q_e}}\right) ,$$

$$G_s = \frac{1}{2}(1-q)\hat{q} + R\hat{R}$$

$$\quad - \int Du \; \ln\left[2\cosh\left(\hat{R} + \sqrt{\hat{q}} u\right)\right] ,$$

$$V(x) = \ln\left[e^\beta + \left(1 - e^{-\beta}\right) H(x)\right] , \quad (3.2.1)$$

where $R_e$ is given above and $q_e = \frac{2}{\pi}\arcsin q$. The value of $q$ is the typical correlation between two $\vec{w}_k^{(0)}$ which are assumed to be independent of the perceptron index $k$. The interpretations of $R$, $\gamma$, $\delta$ and $\alpha$ are as given above. By using the equations generated by the extremal condition in (3.2.1) to eliminate $\hat{R}$ and $\hat{q}$ we can find the dependence of $R$ (and $\epsilon_g$) on $\alpha$ given $\gamma$, $\delta$ and $\beta$.

For $T = 0$ one should, as for the perceptron, find a critical capacity, $\alpha_c$, beyond which the student can not perform optimally on the training set. However, this is not the case implying that the RS-approximation is bad. In the realizable case the values of both the transition and the spinodal point agree with [2]

At $T > 0$ the behavior is much the same as in the high temperature limit with the exception that for $\delta = 1$ the transition is not to an optimal state but to a state approaching optimal learning as $\alpha$ tends to infinity. The asymptotical form of $\epsilon_g$ and $q$ for large $\alpha$ can be found for $\delta = 1$, $\gamma < 1$,

$$\epsilon_g - \epsilon_{opt} = B_1(\gamma,\beta) \left(\frac{\ln\alpha}{\alpha}\right)^2 , \quad (3.2.2)$$

$$1 - q = B(\gamma,\beta) \left(\frac{\ln\alpha}{\alpha}\right)^4 , \quad (3.2.3)$$

and for $\delta < 1$, $\gamma \leq 1$,

$$\epsilon_g - \epsilon_{opt} = B_3(\gamma,\delta,\beta) \left(\frac{\ln\alpha}{\alpha}\right)^{1/3} , \quad (3.2.4)$$

$$1 - q = B_4(\gamma,\delta,\beta) \left(\frac{\ln\alpha}{\alpha}\right)^{4/3} . \quad (3.2.5)$$

The asymptotic behavior can be found by the same method used in appendix B. As for the perceptron there is a range of noise levels for which there is no phase transition at low temperature but one is developed as the temperature is increased. One such example ($\gamma = 1$, $\delta = 0.99$) is used in the phase diagram (7). If the noise is increased above some value there seems to be no phase transition no matter how high the temperature.

## 3.3. Replica Symmetric Breaking.

As was said in the previous section the RS-approximation fails in predicting a critical capacity. Also, the entropy will turn negative at some finite $\alpha$ and thus RSB is expected. In figure (7) a phase diagram for $\gamma = 1$, $\delta = 0.99$ shows the RSB region. For one step of RSB, in the limit $q_1 \to 1$, $\hat{q}_1 \to \infty$, the free energy is,

$$\beta f_{RSB} = \underset{R,\hat{R},q_0,\hat{q}_0,m}{extr} \left[ G_r(R,q_0,m,\alpha,\gamma,\delta,\beta) \right.$$

$$\left. + G_s(R,\hat{R},q_0,\hat{q}_0,m) \right] ,$$

$$G_r = -\frac{2\alpha}{m} \int Du \; H\left[\frac{R_e u}{\sqrt{q_e - R_e^2}}\right] V\left(u\sqrt{\frac{q_e}{1-q_e}}\right) ,$$

$$G_s = \frac{m}{2}(1-q_0)\hat{q}_0 + R\hat{R}$$

$$\quad - \frac{1}{m} \int Du \; \ln\left[2\cosh\left(m\hat{R} + m\sqrt{\hat{q}} u\right)\right] ,$$

$$V(x) = \ln\left[e^{\beta m} + \left(1 - e^{-\beta m}\right) H(x)\right] , \quad (3.3.1)$$

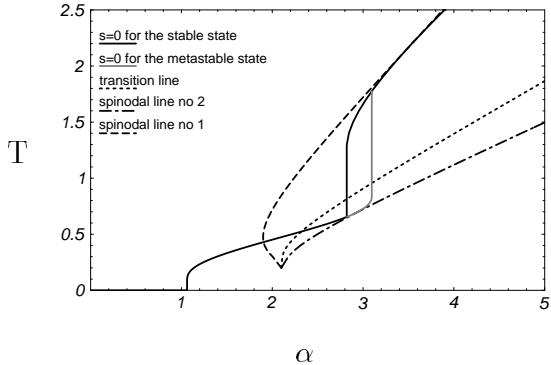

Figure 7: *Phase diagram for the tree committee machine at $\delta = 0.99$, $\gamma = 1.0$. Below the solid line the stable state exhibits RSB and below the gray line the metastable line exhibits RSB. To the left of the first spinodal line there is only one state in the interior. Below the second spinodal line there is only the state close to $R = 1$. In between the two states coexist. Their global properties changes as the transition line is crossed.*

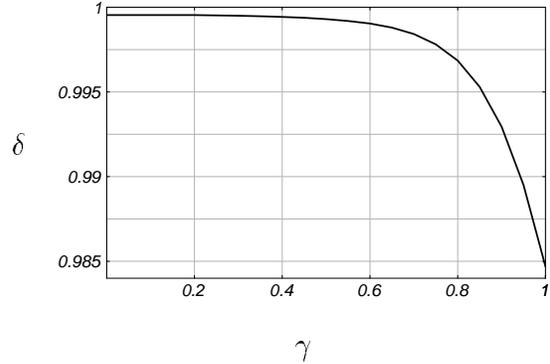

Figure 8: *In the RSB-case of the tree committee machine the system undergoes a phase transition at some $\alpha_{tr}$ at noise levels above the curve. Below it the system will smoothly approach $\epsilon_{opt}$ as $\alpha \to \infty$.*

For reasons analogous to those given in [1] for the perceptron the stationary points of $f_{RSB}$ are given by the stationary points of $f_{RS}$ having zero entropy.

In contrast to the RS-case, but analogous to the high temperature limit, a transition to optimal learning is found for all $\gamma$ if $\delta = 1$. Using the same notation as in section (3.1) the values of $\delta_A$ and $\delta_B$ are 0.9995 and 0.9847 respectively, and $\delta_0(\gamma)$ is given in figure (8).

For the case $\gamma = \delta = 0.05$ the critical capacity, $\alpha_c = 0.95$ ($q_c = 0.31$) is found which is compatible with known results for $\gamma = \delta = 0$ [12]. Typically $\alpha_{sp1}$, $\alpha_{tr}$ and $\alpha_{sp2}$ increase with increasing unrealizability until $\delta = \delta_0(\gamma)$ where $\alpha_{sp1} = \alpha_{tr} = \alpha_{sp2}$. Some values of $\alpha_{sp1}$, $\alpha_{tr}$ and $\alpha_{sp2}$ are given in in table (2), and some typical learning curves are given in figures (9) and (10).

As $\alpha \to \infty$ the asymptotic forms of $\epsilon_g$ and $q$ are,

$$\epsilon_g - \epsilon_{opt} = A_1(\gamma, \delta) \left(\frac{\ln \alpha}{\alpha}\right)^{2/3}, \quad (3.3.2)$$

$$1 - q = A_(\gamma, \delta) \left(\frac{\ln \alpha}{\alpha}\right)^{4/3}. \quad (3.3.3)$$

Appendix B gives details of how to compute the asymptotic behavior, using the perceptron as an example. In the realizable limit ($\delta = \gamma = 1$) the re-sult $\alpha_{sp} = 2.79$, correcting the value found in [2] ($\alpha_{sp} = 2.58$). The reason for the correction is given in appendix A, where the perceptron is used as an example.

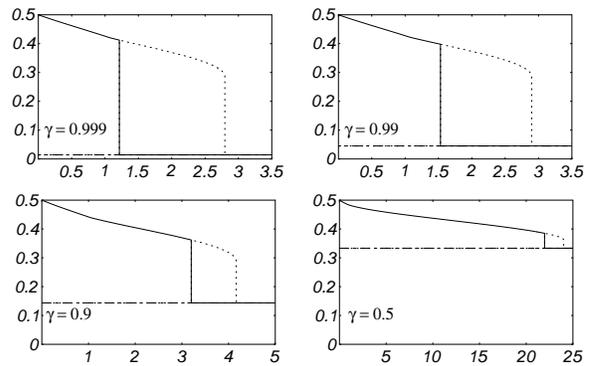

Figure 9: *Learning curves ($\epsilon_g$ vs $\alpha$) for the tree committee machine at $\delta = 1$ and at the indicated $\gamma$. The solid line is the stable state, the dotted the metastable state, and the lowest line in each graph indicates optimal performance.*

## 4. Summary.

In summary we have studied unrealizable learning in two large feed-forward neural network, a perceptron and a tree committee machine within the replica symmetric ansatz as well as for one step of replica symmetry breaking. The average generalization error has been calculated as a function of the load parameter $\alpha$.

For the perceptron it was shown that using a noisy training set results in a generalization error approaching optimal learning with increasing $\alpha$ according to a power law of $(\ln \alpha / \alpha)^k$ with $k = 2$ in the RSB-case. If the noise is low enough there is a phase transition at some finite $\alpha$ to a state which is close to $R = 1$. Increasing the noise makes the transition go away.

For the tree committee machine a similar generalization behavior was found, the main difference being that there is always a transition to optimal learning at some finite $\alpha$ if there is no noise in the input layer. Typically, noise in the input layer gives worse generalization behavior than noise in the hidden layer. For one step of RSB and with noise in the input layer as well as in the hidden layer the asymptotic form of $\epsilon_g$ was found to be $(\ln \alpha / \alpha)^k$ with $k = 2/3$.

In the realizable cases the values of $\alpha_{sp}$ correct previously reported results [1],[2], for the RSB spinodal point in the two machines. Here $\alpha_{sp} = 1.67$ was found for the perceptron and $\alpha_{sp} = 2.79$ for the tree committee machine.

I thank J. Hertz for his valuable advice and direction and R. Urbanczik for many useful discussions. Also, I would like to thank H. Schwarze for sharing the code written in connection to ref. [2] which made it possible to sort out the disagreement on the spinodal points.

## A. The Saddle Point Equations

In the limit $q_1 \to 1$, $\hat{q}_1 \to \infty$ the one step RSB free energy (2.3.1), of the perceptron, is related to the RS-estimate (2.2.1) thru [1],

$$f_{RSB}(R, \hat{R}, q_0, \hat{q}_0, m, \beta)$$
$$= \frac{1}{m} f_{RS}(R, m\hat{R}, q_0, m^2 \hat{q}_0, \beta m). \quad (A.1)$$

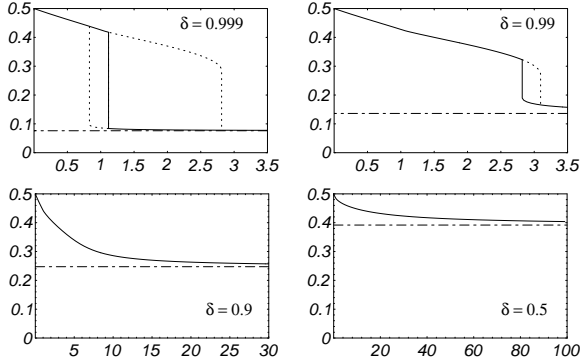

Figure 10: *Learning curves ($\epsilon_g$ vs $\alpha$) for the tree committee machine at $\gamma = 1$ and at the indicated $\delta$. The solid line is the stable state, the dotted the metastable state, and the dashed-dotted line indicates optimal performance.*

| $\delta$ | $\gamma$ | $\alpha_{sp1}$ | $\alpha_{tr}$ | $\alpha_{sp2}$ |
|---|---|---|---|---|
| 1.0000 | 1.0000 | 0.00 | 1.06 | 2.79 |
| | 0.9900 | 0.00 | 1.53 | 2.90 |
| | 0.9000 | 0.00 | 3.20 | 4.16 |
| | 0.5000 | 0.00 | 21.97 | 24.02 |
| 0.9999 | 1.0000 | 0.20 | 1.07 | 2.79 |
| | 0.9900 | 0.36 | 1.07 | 2.90 |
| | 0.9000 | 1.84 | 1.84 | 4.17 |
| | 0.5000 | 20.35 | 20.35 | 24.10 |
| 0.9990 | 1.0000 | 0.83 | 1.12 | 2.81 |
| | 0.9900 | 1.03 | 1.14 | 2.93 |
| | 0.9000 | 2.96 | 2.96 | 4.21 |
| 0.9900 | 1.0000 | 2.82 | 2.82 | 3.10 |
| | 0.9900 | 3.03 | 3.03 | 3.23 |

Table 2: *For one step of RSB for the tree committee machine the table shows some values of $\alpha_{tr}$, $\alpha_{sp1}$ and $\alpha_{sp2}$ for different $\gamma$ and $\delta$*

Stationarity with respect to $R$, $\hat{R}$, $q_0$ and $\hat{q}_0$ results in the relations $q_0(T_{RSB}, m, \alpha) = q_{RS}(T_{RSB}/m, \alpha)$ and $R(T_{RSB}, m, \alpha) = R_{RS}(T_{RSB}/m, \alpha)$ while stationarity with respect to $m$ gives $s_{RS}(T_{RSB}/m, \alpha) = 0$ where $s_{RS}$ is the RS entropy. Thus one can find the stationary points of $f_{RSB}$ by finding stationary points of $f_{RS}$ at a temperature $T_{RS} = T_{RSB}/m$ for which the entropy is zero. The saddle point equations generated by (2.2.1) are

$$q = \int Dt \; \tanh^2(\hat{R} + \sqrt{\hat{q}}t) \;, \quad (A.2)$$

$$R = \int Dt \; \tanh(\hat{R} + \sqrt{\hat{q}}t) \;, \quad (A.3)$$

$$\hat{q} = \frac{\alpha}{\pi(1-q)} \int Du$$
$$\times H\left[\frac{\gamma R u}{\sqrt{q - \gamma^2 R^2}}\right] \frac{e^{-v^2}}{[u_\beta + H(v)]^2} \;, \quad (A.4)$$

$$\hat{R} = \frac{\alpha}{\pi\sqrt{1-q}} \int Dt \; \frac{e^{-y^2/2}}{u_\beta + H(y)} \;, \quad (A.5)$$

with $u_\beta = 1/(e^\beta - 1)$, $v = u\sqrt{q/(1-q)}$ and $y = t\sqrt{(q - \gamma^2 R^2)/(1-q)}$. Using (A.2) and (A.3) to eliminate $q$ and $R$ in (A.4) an (A.5) gives a system of two non-linear equations

$$\hat{q} = \alpha \; h(\hat{R}, \hat{q}) \;, \quad (A.6)$$

$$\hat{R} = \alpha \; g(\hat{R}, \hat{q}) \;. \quad (A.7)$$

At this point we could try to solve for $\hat{q}$ and $\hat{R}$ given $\alpha$. However since $\epsilon_g$ is a many-valued function of $\alpha$ it is more economical to eliminate $\alpha$. This will give the equation,

$$\hat{q} \; g(\hat{R}, \hat{q}) = \hat{R} \; h(\hat{R}, \hat{q}) \;, \quad (A.8)$$

which can be solved for $\hat{q}$ given $\hat{R}$. $\alpha$ can be evaluated using (A.6) or (A.7). The advantage is that $\epsilon_g$ is a single valued function of $\hat{R}$. In the RSB-case this will be helpful since more than one solution for each $\alpha$ has to be considered as we show below.

Once a stationary point has been found its second order properties has to be checked by computing the determinant of the Hessian matrix, $H$. Assume that the correct sign of $\det H$, at $\hat{R} > 0$, is given by the sign at $\hat{R} = 0$. As $\hat{R}$ is increased, the sign of $\det H$ will change first at $\hat{R}_{sp2}$ and then again at $\hat{R}_{sp1}$. Note that $\hat{R}_{sp2} < \hat{R}_{sp1}$ whereas $\alpha_{sp1} < \alpha_{sp2}$. In the regime $\hat{R}_{sp2} < \hat{R} < \hat{R}_{sp1}$, $\beta f$ has a stationary point but it has the incorrect curvature.

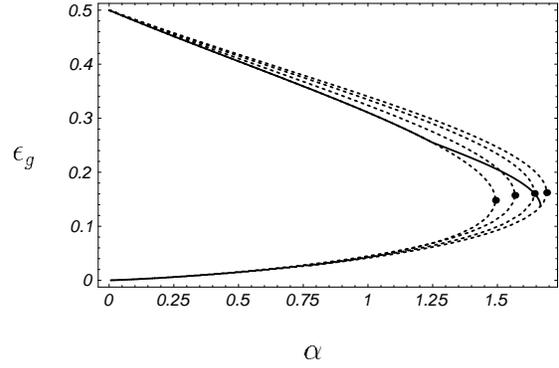

Figure 11: *The dashed learning curves are computed assuming RS for the perceptron at $T = 0$, 0.265, 0.330 and 0.365 reading from top to bottom. The solid learning curve is the one step RSB solution. At the intersection between solid and dashed the entropy for the RS solution is zero. $\hat{R}$ is increased as a RS solution is followed from the top left corner. At the dots $\det H_{RS}$ changes sign. $\det H_{RSB}$ has the same sign every where on the RSB line and is zero at its right end.*

Even though the RSB-case is solved by using the RS-equations the determinant of the Hessian matrix of $\beta f_{RSB}$, $\det H_{RSB}$, has to be used to determine the second order properties. $\det H_{RSB}$ consists of the second derivatives of $\beta f_{RSB}$ with respect to $R$, $q$, $\hat{R}$, $\hat{q}$ and $m$ whereas $\det H_{RS}$ is computed from the second derivatives of $\beta f_{RS}$ with respect to $R$, $q$, $\hat{R}$ and $\hat{q}$. Using $\det H_{RS} = 0$ as the criterion to determine the spinodal point (at $\gamma = 1$) would result in the values of $\alpha_{sp}$ as given in [1] and [2]. Moreover, insisting on this RS-criterion will, for some $\gamma$, result in regions of $\alpha$ where no solution exist. Thus this procedure fails in a disastrous way. However the correct condition, $\det H_{RSB} = 0$, will cure this problem and give $\alpha_{sp}$ as given in this paper.

In the RSB-case $\det H_{RS}$ will have the wrong sign close to $\alpha_{sp}$. This will correspond to points on $\epsilon_g$ not considered in the RS-case since there exists another solution (with $s > 0$) at the same $\alpha$ but with $\det H_{RS}$ having the correct sign. This is illustrated (for $\gamma = 1$) in figure (11).

When the RS-equations are used to solve the RSB-case it is not possible to find the value of $m$ (only $T_{RS} = T_{RSB}/m$). Since $\det H_{RSB}$ depends on $m$ it can not be computed. However it is possible to show that,

$$\det H_{RSB}(R, \hat{R}, q_0, \hat{q}_0, m) = \frac{1}{m} F(R, \hat{R}, q_0, \hat{q}_0) \quad ,\text{(A.9)}$$

and since $0 < m < 1$, $\det H_{RSB}$ has the same sign as $F$.

## B. Asymptotics

For large $\alpha$ the saddle point equations (A.2) implies that $q, R$ are close to 1 and $\hat{q}, \hat{R}$ are large. From this the asymptotic form of the free energy (2.2.1) for non-zero temperatures can be found,

$$\beta f_{RS} = \underset{R,\hat{R},q,\hat{q}}{extr} \left[ G_r(R, q, \alpha, \gamma, \beta) + G_s(R, \hat{R}, q, \hat{q}) \right], \text{(B.1)}$$

$$G_s = \frac{1}{2}(1-q)\hat{q} + R\hat{R} - \sqrt{\frac{2}{\pi}}\sqrt{\hat{q}} \exp\left(-\frac{\hat{R}^2}{2\hat{q}}\right)$$
$$- \hat{R}\left[1 - 2H\left(\frac{\hat{R}}{\sqrt{\hat{q}}}\right)\right] \quad , \tag{B.2}$$

$$G_r = \frac{\alpha\beta}{\pi}\arctan\left[\frac{\sqrt{q - \gamma^2 R^2}}{\gamma R}\right]$$
$$- \frac{\alpha[\phi(\beta) + \phi(-\beta)]}{\sqrt{2\pi}}\sqrt{1-q} \quad , \tag{B.3}$$

$$H(x) = \int_x^\infty \frac{dy}{\sqrt{2\pi}} e^{-\frac{1}{2}y^2} , \tag{B.4}$$

$$\phi(\beta) = \int_0^\infty dw \ln\left[1 + (e^\beta - 1)H(w)\right] . \tag{B.5}$$

If $\gamma < 1$ (B.1) generates the saddle point equations,

$$1 - R = \sqrt{\frac{2}{\pi}}\sqrt{\frac{\hat{q}}{\hat{R}^2}}\exp\left(-\frac{\hat{R}^2}{2\hat{q}}\right) , \tag{B.6}$$

$$1 - q = \sqrt{\frac{2}{\pi}}\frac{1}{\sqrt{\hat{q}}}\exp\left(-\frac{\hat{R}^2}{2\hat{q}}\right) , \tag{B.7}$$

$$\hat{q} = \frac{\alpha\beta\gamma}{\pi\sqrt{1-\gamma^2}} + \frac{\alpha[\phi(\beta) + \phi(-\beta)]}{\sqrt{2\pi}\sqrt{1-q}} , \tag{B.8}$$

$$\hat{R} = \frac{\alpha\beta\gamma}{\pi\sqrt{1-\gamma^2}} . \tag{B.9}$$

The first two of these can be combined into,

$$\frac{1-R}{1-q} = \frac{\hat{q}}{\hat{R}} . \tag{B.10}$$

Using (B.6) and (B.8)-(B.10) results in,

$$1 - q \sim (1 - R)^2 , \tag{B.11}$$

$$(1 - R)^{3/2} \sim \frac{1}{\sqrt{\alpha}}\exp\left(-\alpha A_2(1-R)\right) , \tag{B.12}$$

where $A_2$ depend only on $\beta$ and $\gamma$ and where $\sim$ means proportional to in the asymptotic limit of large $\alpha$. In order to solve (B.12) the ansatz,

$$1 - R(\alpha) = A_0\frac{\ln\alpha}{\alpha} + \delta(\alpha) , \tag{B.13}$$

is made. For consistency it is important to check that $\delta(\alpha)$ is of lower order than $\ln(\alpha)/\alpha$. Combining (B.12) with the ansatz (B.13) and by choosing $A_0 = 1/A_2$ gives the solution,

$$\epsilon_g - \epsilon_{opt} \sim 1 - R \sim \frac{\ln\alpha}{\alpha} , \tag{B.14}$$

Also $\delta(\alpha) \sim \ln[\ln(A_0 \ln\alpha)/A_2^{2/3}]/\alpha$ is found and thus $\delta(\alpha)$ is of lower order. The asymptotic form of $1 - q$ is now easily found using (B.14) and (B.11).

In the RSB-case the temperature is given by the zero entropy condition and can not be regarded as an arbitrary constant. Thus $\beta$ is a function of $\alpha$ and combining the saddle point equations (B.6)-(B.10) with the asymptotic form of the zero entropy condition, $\hat{q} \sim (1-q)^{-1/2}$, gives,

$$1 - q \sim \beta^2 , \tag{B.15}$$
$$1 - R \sim \beta^2 , \tag{B.16}$$
$$\beta^{5/2} \sim \frac{1}{\sqrt{\alpha}}\exp(-B_2\alpha\beta) , \tag{B.17}$$
$$\tag{B.18}$$

where $B_2$ only depend on $\gamma$. Again an ansatz, $\beta(\alpha) = B_0 \ln(\alpha)/\alpha + \delta(\alpha)$ is made which together with $B_0 = 2/B_2$ gives the solution,

$$\beta(\alpha) \sim \frac{\ln\alpha}{\alpha} . \tag{B.19}$$

The asymptotic form of $R$, $q$ and $\epsilon_g$ is now found from (B.15) and (B.16) giving,

$$\epsilon_g - \epsilon_{opt} \sim 1 - R \sim \left(\frac{\ln\alpha}{\alpha}\right)^2 , \tag{B.20}$$

For the tree committee machine the asymptotic forms of $R$, $q$ and $\epsilon_g$ can be found by the same procedure but using the asymptotic form of the free energy (3.2.1).

# References

[1] H. S. Seung, H. Sompolinsky, and N. Tishby, *Phys. Rev. A*, **45** 6056(1992).

[2] H. Schwarze and J. Hertz, *Europhys. Lett.*, **20** 375(1992).

[3] J. Hertz, A. Krogh, and R.G. Palmer, *Introduction to the theory of Neural Computation* (Addison-Weslay, Redwood City, CA, 1991).

[4] M. Mézard, G. Parisi, and M. A. Virasoro, *Spin Glass Theory and Beyond* (World Scientific, Singapore, 1987).

[5] T. L. H. Watkin, A. Rau, and M. Biehl, *Rev. Mod. Phys.*, **65** 499(1993).

[6] G. Györgyi and N. Tishby, in *Neural Networks and Spin Glasses*, edited by W. K. Theumann and R. Köberle. (World Scientific, Singapore, 1990).

[7] R. Urbanczik, *A large committee machine learning unrealizable rules*, submitted to *Neural Computation*, 1995.

[8] R. Urbanczik, *A fully connected committee machine learning unrealizable rules*, to be published.

[9] E. Gardner, *J. Phys. A*, **21** 271(1988).

[10] W. Krauth and M. Mézard, *J. Phys. Paris*, **50** 3057(1989).

[11] D. Haussler, M. Kearns, H. S. Seung and, N. Tishby, *Rigorous Learning Curve Bounds from Statistical Mechanics*, to be published.

[12] E. Barkai, D. Hansel, and H. Sompolinsky, *Phys. Rev. A*, **45** 4146(1992).